\documentstyle[aps,epsf]{revtex}  % DO NOT DELETE THIS LINE 

\begin{document}        %  DO NOT DELETE OR CHANGE THIS LINE

\baselineskip 14pt
\title{Wormholes and Flux Tubes in Kaluza-Klein Theory}
\author{Vladimir Dzhunushaliev}
\address{Theor. Phys. Dept. KSNU, 720024, Bishkek, Kyrgyzstan and
Universit{\"a}t Potsdam, Institut f{\"u}r Mathematik, 14469 Potsdam,
Germany}
\author{Douglas Singleton}   % Use this and the next line only if there is a second
\address{Dept. of Phys. CSU Fresno, 2345 East San Ramon Ave. M/S 37
Fresno, CA 93740-8031 USA}  % address. (Remove the left % marks)
\maketitle              % Creates the title area, Do Not Remove

\begin{abstract}        % Do Not Delete this line
In this work spherically symmetric solutions to 5D Kaluza-Klein
theory, with ``electric'' and/or ``magnetic'' fields are examined.
Different relative strengths of the ``electric'' and ``magnetic''
charges of the solutions are studied by varying certain parameters
in our metric ansatz. As the strengths of these two charges
are varied the resultant spacetime exhibits an interesting ``evolution''.
\
\end{abstract}   	% Do Not Delete this line

\section{Introduction}               % Introduction goes below.

In this work we investigate a class of spherically symmetric
metrics in multidimensional (MD) gravity. The metric ansatz which
is used has off diagonal elements which leads to these solutions
having ``electric'' and/or ``magnetic'' charges.
The solutions examined here are either MD wormholes or
infinite/finite flux tubes. It is found that the type of
solution obtained depends crucially on the relative
magnitudes of these charges and thus on the
form of the off-diagonal metric components. Usually in the
discussion of wormhole or blackhole solutions such off-diagonal elements
are not considered, (see for, example, \cite{mor}-\cite{vis}). 
\par 
The off-diagonal components of the MD metric play the role of 
gauge fields (U(1), SU(2) or SU(3) gauge fields), and 
a scalar field $\phi(x^\mu)$ which is connected with the 
linear size of the extra dimension. 
These geometrical fields can act as the source of the 
exotic matter necessary for the formation of
the wormhole's mouth. Such solutions were obtained in Refs.
\cite{chodos} \cite{clem} \cite{dzh1} \cite{dzh2}.
These works indicate that the exotic matter necessary for the
formation of the WH can appear in \textbf{\textit{vacuum multidimensional 
gravity}} from the off-diagonal elements of the metric
(the gauge fields) and from the $G_{55}$ component of the metric
(the scalar field), rather than coming from some externally
given exotic matter.
\par 
In Refs. \cite{dzh1}, \cite{dzh2} a MD metric with
only ``electric'' fields was investigated. In Ref \cite{dzh4}
a MD metric with  ``magnetic'' field = ``electrical'' field  
was investigated.
In this paper we investigate the consequence of having
both ``electric'' and  ``magnetic'' Kaluza-Klein fields of
varying relative strengths. We will consider 5D 
Kaluza-Klein theory as gravity on the principal 
bundle with U(1) fibre and 4D space as the base of this bundle 
\cite{dzh2}. 

\section{The field equations}

For our spherically symmetric 5D metric we take 
\begin{eqnarray} 
ds^2 &=& e^{2\nu (r)}dt^{2} - r_0^2e^{2\psi (r) - 2\nu (r)}
\left [d\chi +  \omega (r)dt + n\cos \theta d\varphi \right ]^2  
\nonumber \\
&-& dr^{2} - a(r)(d\theta ^{2} + 
\sin ^{2}\theta  d\varphi ^2),
\label{1}
\end{eqnarray}
where $\chi $ is the 5$^{th}$ extra coordinate; 
$r,\theta ,\varphi$ are $3D$  spherical-polar coordinates; 
$n$ is integer; $r \in \{ -R_0 , +R_0 \}$ 
($R_0$ may be equal to $\infty$). We require that all 
functions $\nu (r), \psi(r)$ and $a(r)$ be even 
functions of $r$ and hence  $\nu'(0)=\psi'(0)=a'(0)=0$.
The ansatz function $\omega (r)$ is the $t$-component of the electromagnetic
potential and $(n\cos\theta)$ is the $\varphi$-component. 
Thus we have radial Kaluza-Klein ``electrical'' and ``magnetic'' fields. 
\par 
Substituting this ansatz into the 5D Einstein vacuum equations 
gives \cite{dzhds} the following set of coupled, non-linear
differential equations
\begin{eqnarray} 
\nu '' + \nu'\psi' + \frac{a'\nu'}{a} - 
\frac{1}{2} r_0^2 \omega '^2e^{2\psi - 4\nu} = 0,
\label{3}\\
\omega '' - 4\nu'\omega' + 3\omega '\psi ' + 
\frac{a'\omega '}{a} = 0, 
\label{4}\\ 
\frac{a''}{a} + \frac{a'\psi '}{a} - \frac{2}{a} + 
\frac{Q^2}{a^2}e^{2\psi - 2\nu} = 0, 
\label{5}\\ 
\psi '' + {\psi '}^2 + \frac{a'\psi '}{a} - 
\frac{Q^2}{2a^2}e^{2\psi - 2\nu} = 0, 
\label{6}\\ 
\nu '^2 - \nu '\psi ' - \frac{a'\psi '}{a} + 
\frac{1}{a} - \frac{a'^2}{4a^2} - 
\frac{1}{4}r_0^2\omega '^2 e^{2\psi - 4\nu} - 
\frac{Q^2}{4a^2} e^{2\psi - 2\nu} = 0 
\label{7} 
\end{eqnarray} 
The Kaluza-Klein ``magnetic'' charge is $Q = nr_0$.
The Kaluza-Klein ``electrical'' field can be defined
by multiplying Eq. (\ref{4}) by $4 \pi r_0$ 
and rewriting it as
\begin{equation} 
\left( r_0 \omega ' e^{3\psi - 4\nu} 4 \pi a \right)' = 0.
\label{7a} 
\end{equation} 
This can be compared with the normal 4D Gauss's Law 
\begin{equation} 
\left ( E_{4D} S\right )' = 0,
\label{7b}
\end{equation} 
where $E_{4D}$ is 4D electrical field and $S = 4 \pi r^2$ is the 
area of 2-sphere $S^2$. Eqs. (\ref{3}) - (\ref{7}) are five equations for
determining the four ansatz functions ($\nu, \psi ,
a , \omega$). The first four equations (Eqs. (\ref{3} - \ref{6}))
are dynamical equations which determine the ansatz functions, 
while the last equation (Eq. (\ref{7})) contains no new dynamical
information not contained in the first four equations, but
gives some initial conditions related to solving this system
of equations. For the metric given in Eq. (\ref{1})
$r^2$ is replaced by $a(r)$ and the surface area 
is given by $S = 4 \pi a (r)$. Comparing Eq. (\ref{7a})
with Eq. (\ref{7b}) we can identify the 5D Kaluza-Klein
``electric'' field as
\begin{equation}
\label{7c}
E_{KK} = r_0 \omega ' e^{3 \psi - 4 \nu}
\end{equation}
If we integrate Eq. (\ref{7a}) once and let the integration
constant be $4 \pi q$, then from Eq. (\ref{7c}) we
find that $E_{KK} = q / a(r)$ where $q$ can be taken as the
Kaluza-Klein ``electric'' charge.
Finally for the system of equations given in Eqs.
(\ref{3} ) - ( \ref{7} ) we will consider solutions with
the boundary conditions $a(0) = 1 , \psi (0) = \nu (0) = 0$ 
(for numerical calculations we introduced dimensionless 
function $a(r)\to a(r)/a(0)$ and $x = r/a(0)$).
Using these boundary conditions in Eq. (\ref{7}) and also
in Eq. (\ref{7c}) (which gives $r _0 w' (0) = q$) gives the
following relationship between the Kaluza-Klein ``electric''
and ``magnetic'' charges
\begin{equation}
\label{7d}
1 = {q^2 + Q^2 \over 4a(0)}
\end{equation}
From Eq. (\ref{7d}) it is seen that the charges can be parameterized
as $q = 2 \sqrt{a(0)}\sin \alpha$ and $Q = 2 \sqrt{a(0)}\cos \alpha$.
\par 
We will examine the following cases: 
\begin{description} 
\item[A)] 
$Q=0$ or $H_{KK} = 0$ , ``magnetic'' field is zero.
\item[B)]
$q=0$ or $E_{KK} = 0$ , ``electrical'' field is zero. 
\item [C)]
$H_{KK} = E_{KK}$, ``electrical'' field equal to ``magnetic'' 
field.  
\item[D)] 
$H_{KK} < E_{KK}$, ``magnetic'' field less 
than ``electrical''. 
\item[E)] 
$H_{KK} > E_{KK}$, ``electrical'' field less 
than ``magnetic''.
\end{description}

\subsection{Switched off ``magnetic'' field.} 

In this case we have the following solution \cite{chodos} \cite{dzh1}: 
\begin{eqnarray}
a & = & r^{2}_{0} + r^{2},
\label{8}\\
e^{2\nu } & = & {2r_{0}\over q}{r^{2}_{0}+r^{2}
\over r^{2}_{0}-r^{2}},
\label{9}\\
\psi &=& 0
\label{9a}\\
\omega & = & 4r_{0}\over q}
{r\over {r^{2}_{0} - r^{2}}.
\label{10}
\end{eqnarray}
This WH-like spacetime has a nonasymptotically flat metric, 
bounded by two surfaces at $r=\pm r_0$ where the
reduction from 5D to 4D spacetime breaks down. As $r$ moves
away from $0$ the cross-sectional size of the throat, $a(r)$, 
increases. 

A connection can be made between the present solution and 
Wheeler's old proposal of electric charge as a wormhole filled
with electric flux that flows from one mouth to the other --
the ``charge without charge'' model of electric charge. 
In a recent work \cite{dzh3} a model of electric charge 
along these lines was proposed where electric charge is 
modeled as a kind of composite WH with a 
quantum mechanical splitting off of the 5$^{th}$ dimension. 
The 5D WH-like solution of Eqs. (\ref{8}-\ref{10}) 
have two Reissner-Nordstr\"om black holes attached to
it on the surfaces at $\pm r_0$. By considering 4D electrogravity
as a 5D Kaluza-Klein theory in the initial Kaluza formulation 
with $G_{55}=1$ we can join the 5D and Reissner-Nordstr\"om 
solutions at the $r=\pm r_0$ surfaces base to base and 
fibre to fibre.

\subsection{Switched off ``electrical'' field}

In this case we will simplify by taking $\nu = 0$ in addition to
$\omega =0$ so that the equations reduce to
\begin{eqnarray} 
\frac{y''}{y} + \frac{y'a'}{ya} - \frac{Q^2 y^2}{2a^2} = 0, 
\label{11} \\
\frac{a''}{a} + \frac{y'a'}{ya} - \frac{2}{a} + 
\frac{Q^2 y^2}{a^2} = 0, 
\label{12} \\
\frac{a'y'}{ay} - \frac{1}{a} + \frac{a'^2}{4a^2} + 
\frac{Q^2y^2}{4a^2} = 0 
\label{13} 
\end{eqnarray} 
where $y(r)=\exp{(\psi(r))}$. These are three equations for two
ansatz functions, $\psi (r) , a(r)$. The last equation, Eq. (\ref{13}),
simply repeats information that is already contained in the
first two equations. We solved the system of equations 
(\ref{11}) - (\ref{12}) numerically, using the {\it Mathematica}
package, with the following initial conditions: 
$a(0) = a_0 = 1$, $a'(0) = 0$, $y(0) = 1, y'(0) = 0$, 
(where we are using the dimensionless quantities 
$x=r/a_0$ and $a \rightarrow a/a_0$). These conditions
and $\alpha = 0$ fix the dimensionless ``magnetic'' charge as 
$Q =  2$. The detailed results of the numerical calculations 
for $a(r)$ and $y(r)$ are given in Ref. \cite{dzhds}. The
general shape of the cross-section function, $a(r)$ of this
solution can be seen in the last picture of Fig. 1.
Also from this picture one can see two singularities at $x= \pm x_0$ .
We interpret these singularities as the 
location of two magnetic charges ($\pm Q$) 
with flux lines of Kaluza-Klein ``magnetic'' 
field going from $+Q$ to $-Q$.
Near these singularities the ansatz
functions have the following asymptotic behaviour: 
\begin{eqnarray} 
y(r) \approx \frac{y_{\infty}}{(r_0 - r)^{1/3}}, 
\label{14} \\
a(r) \approx a_\infty (r_0 - r)^{2/3}, 
\label{15} \\
\frac{Q y_\infty}{a_\infty} = \frac{2}{3}. 
\label{16} 
\end{eqnarray} 
The time part of the metric appears 
not to be influenced by the strong gravitational field since 
$G_{tt}(r) = \exp{(2\nu (r))} = 1$. This result is similar
to what was found in Ref. \cite{gross} \cite{sorkin} where ``magnetic''  
Kaluza-Klein components of the metric were considered. One difference
between the present solutions and the monopole solutions of Ref.
\cite{gross} \cite{sorkin}, is that the monopole solutions had
only coordinate singularities, while $r=\pm r_0$ are real
singularities for the present solution. This can be seen by 
calculating the invariant $R_{AB} R^{AB}$ and using the asymptotic
form for $y(r), a(r)$ given in Eqs. (\ref{14}) - (\ref{16}). In
Ref. \cite{dzhds} it was found that
\begin{equation}
R_{AB} R^{AB} \propto {1 \over (r_0 -r)^2}
\end{equation}
\par 
Finally it can be shown that
this spacetime has a finite volume $V$,
by calculating $V=\int\sqrt{-G}d^5v$. Near the singularities 
$r=\pm r_0$ we have: 
\begin{equation} 
\sqrt{-G} = \sqrt{-det (G_{AB})} = 
r_0 a(r) \exp{(\psi (r)}) \sin\theta 
\approx (r_0 - r)^{1/3} \rightarrow 0 
\label{17}
\end{equation} 
The form of this solution is suggestive of the color field flux 
tubes which are conjectured to form between two quarks 
in some pictures of confinement (see for example pg. 548
of Ref. \cite{peskin}).
\par

\subsection{``Magnetic'' field  equal to ``electrical'' field}

In this case $Q=q$ and an exact solution can be given \cite{dzh4}: 
\begin{eqnarray} 
a = \frac{q^2}{2} = const, 
\label{18}\\
e^{\psi} = e^{\nu} = \cosh\frac{r\sqrt{2}}{q},
\label{19}\\
\omega = \frac{\sqrt{2}}{r_0}\sinh\frac{r\sqrt{2}}{q} 
\label{20}
\end{eqnarray} 
Using this solution and Eq. (\ref{7c}) 
we find that the Kaluza-Klein ``electrical'' field is
\begin{equation} 
E_{KK} = \frac{q}{a} = \frac{2}{q} = const.
\label{23} 
\end{equation}
A similar magnetic flux tube-like solution was discussed in
Ref. \cite{davidson}.
The Kaluza-Klein ``magnetic'' field 
associated with this solution is \cite{dzh4}
\begin{equation}
H_{KK} = \frac{r_0 n}{a} = \frac{Q}{a} = const. 
\label{24}
\end{equation}
Thus, this solution is \textit{\textbf{an infinite flux tube}} 
with constant Kaluza-Klein ``electrical'' and 
``magnetic'' fields. The direction of both the
``electric'' and ``magnetic'' fields is along the
${\hat {\bf r}}$ direction ({\it i.e.} along the axis
of the flux tube). The sources 
of these Kaluza-Klein fields (5D ``electrical'' 
and ``magnetic'' charges) are located at $\pm \infty$. 
This feature leads us to consider this solution as a kind
of 5D  ``electrical'' and ``magnetic'' dipole. 

\subsection{Intermediate cases} 

We consider two different cases: 
$E_{KK} > H_{KK}$ (or $q > Q$) and $E_{KK} < H_{KK}$
(or $q<Q$). The initial conditions  for both cases are 
taken as : $\psi (0) = \nu (0) = 0$, 
$\psi '(0) = \nu '(0) = 0$ and $a(0) = 1, a'(0) = 0$. These
initial conditions along with a choice of $\alpha$ determine
the magnitude of the charges $q , Q$. As in the ``magnetic'' case we solved
the system of equations numerically \cite{dzhds}.

\subsubsection{$E_{KK} > H_{KK}$}

As the ``magnetic'' field increases from $0$ 
to $H_{KK}=E_{KK}$ we found the following behaviour:
First, compared to the WH-like solution
of the pure ``electric'' case, the longitudinal distance 
between the surfaces $\pm r_0$ is stretched as the magnetic field
strength increases; second, the cross-sectional
size of the solution, represented by the function $a(r)$ did not
increase as rapidly as $r \rightarrow \pm r_0$. In the limit
where the ``magnetic'' field equals the ``electrical'' field, 
$H_{KK}=E_{KK}$, the longitudinal length of the solution
goes to $\infty$ and the cross-sectional  size became a
constant.

\subsubsection{$E_{KK} < H_{KK}$}

In this case the ``electrical'' field is taken
as decreasing from the $E_{KK}=H_{KK}$ down to
$E_{KK}=0$. As the ``magnetic'' field strength increases
relative to the ``electric'' field strength we notice
the following evolution of the solution : the infinite
flux tube of the equal field case turns into a finite flux
tube when $E_{KK}$ drops below $H_{KK}$. Also the cross-sectional
size of this case has a maximum at $r=0$ and decreases as
$r \rightarrow \pm r_0$ where the singularities occur. We take these
singularities as the locations of the ``electric'' / ``magnetic''
charges. Between the charges there is a flux tube of
Kaluza-Klein ``electric'' and ``magnetic'' fields. The longitudinal 
size of this flux tube (the distance between charges) reaches its
minimum in the limit when there is only a ``magnetic'' field
($E_{KK} = 0$).

\section{Discussion} 

As the relative strengths of the Kaluza-Klein fields are
varied we find that the solutions to the metric in
Eq. (\ref{1}) evolve in a very interesting and suggestive way :
\begin{enumerate} 
\item 
$0 \le H_{KK} <E_{KK} $. The solution 
is \textbf{\textit{a WH-like object}} 
located between two surfaces at $\pm r_0$ where the
reduction of 5D to 4D spacetime breaks down. The cross-sectional
size of these solution increases as $r$ goes from $0$
to $\pm r_0$. The throat between the $\pm r_0$ surfaces is
filled with ``electric''and/or ``magnetic'' flux. As the strength
of the ``magnetic'' field is increased the longitudinal
distance between the $\pm r_0$ surfaces increases, and
the cross-sectional size does not increase as rapidly
as $r \rightarrow r_0$.
\item 
$H_{KK} = E_{KK}$. In this case the solution is 
\textbf{\textit{an infinite flux tube}} filled
with constant ``electrical'' and ``magnetic'' fields, and
with the charges disposed at $\pm \infty$. The cross-sectional
size of this solution is constant ($ a= const.$). Essentially,
as the magnetic field strength is increased one can think that
the previous solutions are stretched so that the
$\pm r_0$ surfaces are taken to $\pm \infty$ and the cross section
becomes constant.  
\item 
$0 \le E_{KK} < H_{KK}$. In this case we have
\textbf{\textit{a finite flux tube}} located 
between two (+) and (-) ``magnetic'' and/or ``electric'' 
charges located at $\pm r_0$. Thus the longitudinal 
size of this object is again finite, but now the cross
sectional size decreases as $r \rightarrow r_0$. At
$r = \pm r_0$ this solution has real singularities which
we interpret as the locations of the charges. 
This solution is very similar to the confinement 
mechanism in QCD where two quarks are 
disposed at the ends of a flux tube with color electrical 
and magnetic fields running between the quarks.
In the $E_{KK} = 0$ limit we find two opposite ``magnetic'' charges
confined to a spacetime of fixed volume.
This may indicate why single, asymptotic magnetic charges
have never been observed in Nature : they are permanently
confined into monopole-antimonopole pairs of some fixed
volume. Finally,  we note that in Ref. \cite{singl}
some similar mappings between 4D gravity and non-Abelian 
theory are discussed.
\end{enumerate} 
\par 
The evolution of the solution from a WH-like object, to
an infinite flux tube, to a finite flux tube, as the
relative strengths of the fields is varied, is summarized in
Fig.1. This allows two complimentary conclusions :
First, if one takes some Wheeler-like model of electric
charge as in Ref. \cite{dzh3} then it can be seen that
if the magnetic field becomes too strong the WH-like
solution is destroyed and with it the Wheeler-like
model of electric charge.
Second, if one concentrates a sufficently strong electric
field ({\it i.e.} $E_{KK} > H_{KK}$) into some small region 
of spacetime one is led to the science fiction-like 
possibility that one may be able 
\textit{\textbf{to ``open'' the finite
flux tubes into a WH-like configuration}}. This 
conjecture assumes some kind of spacetime foam model
where the vacuum is populated by virtual flux tubes
filled with virtual ``magnetic'' and/or ``electric''
fields.
\par 
Starting from the solutions obtained here we see that in 5D 
gravity there is a distinction between ``electrical'' and 
``magnetic'' Kaluza-Klein fields. This can be contrasted
with the 4D electrogravity Reissner-Nordstr\"om solution 
which is the same for the electrical and magnetic charges. 

\section{Acknowledgements} 

This work has been funded by the
National Research Council under the Collaboration
in Basic Science and Engineering Program.

%% ONLY POSTSCRIPT FILES CAN BE INCLUDED

\begin{figure}[ht]	% in second brace, h=here, t=top, b=bottom	
\centerline{\epsfxsize 3.8 truein \epsfbox{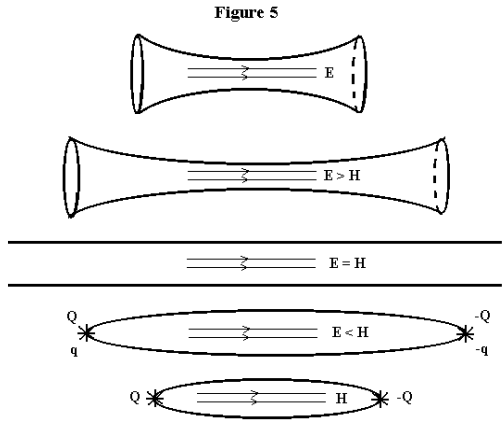}}   
\vskip -3.0 cm
\caption[]{
\label{whfig}
\small The evolution of the ``electric''/``magnetic'' solutions
as a function of the relative strengths between the two charges}
\end{figure}

\end{document}